\documentclass[11pt]{iopjournal}
\usepackage[english]{babel}
\usepackage{amsmath, amssymb, amsfonts}
\usepackage{graphicx}
\usepackage{color}
\usepackage{caption}
\usepackage{subcaption}
\usepackage{hyperref}
\hypersetup{
    colorlinks=true,
    linkcolor=blue,
    citecolor=blue,
    urlcolor=blue
}

\usepackage[
  backend=biber,
  style=numeric-comp,    
  sorting=none,
  doi=true,
  url=false,             
  isbn=false
]{biblatex}

\def\abs#1{\left\vert #1 \right\vert}

\newcommand{\haat}[1]{\expandafter\hat#1}
\begin{document}

\title{Engineering Anderson Localization in Arbitrary Dimensions with Interacting Quasiperiodic Kicked Bosons}

\author{H. Olsen$^{1,*}$, P. Vignolo$^{1,2}$, M. Albert$^{1,2}$}

\affil{$^1$Universit\'e C\^ote d'Azur, CNRS, Institut de Physique de Nice, France}

\affil{$^2$Institut Universitaire de France}

\affil{$^*$Corresponding author: \href{mailto:hazel.olsen@univ-cotedazur.fr}{hazel.olsen@univ-cotedazur.fr}}

\begin{abstract}
   We study the interplay of interactions and quasiperiodic driving in the Lieb–Liniger model of one-dimensional bosons subjected to a sequence of delta kicks. Building on the known mapping between the kicked rotor and the Anderson model, we show that both interparticle interactions and quasiperiodic modulations of the kicking strength can independently and simultaneously generate synthetic dimensions. In the absence of modulation, interactions between two bosons already promote an effective two-dimensional Anderson model. Introducing one or two additional incommensurate frequencies further extends the system to three and four effective dimensions, respectively. Through extensive numerical simulations of the two-body dynamics and finite-time scaling analysis, we observe Anderson localization and the associated critical behavior characteristic of the orthogonal universality class. This combined use of interactions and quasiperiodic driving thus provides a versatile framework for emulating Anderson localization and its transition in arbitrary dimensions. 
\end{abstract}

\section{Introduction\label{sec:intro}}

\medskip

Anderson localization is the complete suppression of wave propagation due to interference effects in disordered media. Although originally introduced in the context of electron transport \cite{Anderson1958}, Anderson localization has since been ubiquitously observed in classical and quantum wave systems \cite{Katsumoto1987,RevModPhys.57.287,Dalichaouch1991,chabanov2000statistical,Schwartz2007,Chabe2008,Billy2008,Roati2008,Hu2008,Lahini2008,Kondov2011,Jendrzejewski2012,Mordechai1997,Yamilov2023}. An interesting aspect of Anderson localization is that, in the absence of interactions, a metal-insulator transition can be observed between localized and delocalized states in three or more dimensions \cite{Thouless1974}. A critical level of disorder separates these two phases, once reached, interference effects mean eigenstates become exponentially localized and diffusion is suppressed. This type of metal-insulator transition is known as the Anderson transition. It can be linked to second-order phase transitions, allowing us to formulate the so-called one parameter scaling theory of localization \cite{Abrahams1979}. This scaling theory tells us that the dynamics is universal and scale invariant in the vicinity of the transition, here the critical behavior can be described by a single scaling variable. When the disorder strength $W$ is less than the critical disorder $W_c$, the localization length diverges at criticality as $\ell\sim (W-W_c)^{-\nu}$, for $\nu$ the critical exponent. On the other side of the transition we obtain the diffusion constant $D$ which vanishes as $(W_c-W)^{s}$, for a new critical exponent $s$. These critical exponents are related by Wegner's scaling law \cite{Wegner1976}, which states 
\begin{equation}
 \label{eq:wegner}
    s = (d - 2) \nu
\end{equation}
with $d$ being the dimensionality of the system. From this relation we know that $s$ and $\nu$ are equivalent in three dimensional systems, but not for systems with four or more dimensions, complicating the extraction of these exponents. Exploring dimensionality is particularly compelling in the quantum regime, where phase transitions are driven by quantum rather than thermal fluctuations. The strength of these fluctuations, and hence the existence and nature of the transition, depend sensitively on dimensionality. Increasing $d$, and with it the number of neighbors, typically enhances the role of average quantities so that mean-field theories become exact above a certain upper critical dimension, which is four for many classical and quantum systems. In contrast, Anderson localization exhibits a strikingly different behavior. Disorder-induced interference effects persist in all dimensions, and numerical as well as analytical studies suggest that no finite upper critical dimension exists for the Anderson transition \cite{Evers2008,Mirlin1994,Slevin2014,Tarzia2017}. This implies that nontrivial critical behavior may survive even in very high dimensions, making it a unique quantum phase transition that defies the usual mean-field paradigm. Investigating how the critical exponents and scaling functions evolve with increasing dimensionality therefore provides a rare opportunity to probe the limits of universality in disordered quantum systems. Another fascinating layer of Anderson localization emerges once interactions are taken into account. The effect of interactions in Anderson localized systems is an open question which is especially intriguing for low-dimensional or driven systems. Typically interactions promote delocalization \cite{GiamarchiSchulz1988}, however when strong disorder is present in a system, typically of low dimensions, a many-body localized phase can be produced \cite{Basko2006, NandkishoreHuse2015, AletLaflorencie2018, Ponte2015a, Ponte2015b}. This phase is not ergodic, and thus prevents thermalization. Understanding how Anderson localisation is effected by both interactions and dimensionality therefore becomes an important task.

One interesting platform for exploring Anderson localization is the quantum kicked rotor \cite{SANTHANAM20221}, a paradigmatic model of both classical and quantum chaos. In its quantum realization, the periodically driven rotor exhibits \emph{dynamical localization}—the direct analog of Anderson localization, but occurring in momentum space. This phenomenon arises from destructive interference between quantum amplitudes, leading to a saturation of the kinetic energy at long times \cite{Fishman1982, Grempel1984, Shepelyansky1986}. Over the past decades, the quantum kicked rotor has been implemented in numerous atomic experiments, establishing it as one of the most versatile systems to probe localization phenomena and quantum transport \cite{Moore1995,Chabe2008,Lemarie2010a,Lopez2012,Manai2015,Hainaut2018,Gupta2021,Cao2022,Nagerl2023,Madani2025}. Among these, ultracold atomic gases—particularly one-dimensional (1D) Bose gases—offer an exceptionally clean and controllable realization \cite{Garreau2008}. Such systems are accurately described by the Lieb–Liniger model \cite{LiebLiniger}, and allow fine-tuned control of the interaction strength through established experimental techniques \cite{Paredesetal, Kinoshita2004, Kinoshita2005, CazalillaCitroGiamarchiOrigancRigol}. This makes them ideal candidates for investigating the interplay between periodic driving, interactions, and localization. These systems are, however, seemingly restricted by their one-dimensional nature. In recent years, various experimental techniques have been developed to investigate the role of dimensionality in dynamically localized systems. To observe an Anderson transition with the quantum kicked rotor, the system must be generalized to one that is equivalent to an Anderson model in three or more dimensions. Several approaches have been proposed to introduce synthetic dimensions into the kicked-rotor framework. One particularly effective route is the quasiperiodic generalization of the kicked rotor, in which the amplitude of the standing-wave pulses driving the system is modulated in time. By carefully choosing incommensurate modulation frequencies, it is possible to increase the effective dimensionality of the system~\cite{Casati1989b, Chabe2008, Toh2024, Madani2025}. Interactions have also been shown to alter the effective dimensionality of the quantum kicked rotor~\cite{Olsen2025, Landini2025,Notarnicola2018}. However, the combined influence of interactions and quasiperiodic driving—an interacting quasiperiodic kicked rotor—has not yet been explored. In this work, we investigate such a system, bridging these two mechanisms to engineer higher-dimensional Anderson models within a simple and controllable setting. We demonstrate that synthetic dimensions can be generated both by introducing interparticle interactions and by modulating the kicking strength with additional incommensurate frequencies, and that both mechanisms can be employed simultaneously. In particular, we show that a minimal model of two interacting bosons can emulate Anderson localization in two, three, and four effective dimensions when subject to zero, one, or two additional driving frequencies, respectively. By analyzing the quantum dynamics and performing finite-time scaling, we identify the corresponding Anderson transitions and extract critical exponents consistent with the orthogonal universality class. This approach can be naturally extended to systems with more particles and additional modulation frequencies, providing a versatile platform for studying Anderson localization in arbitrary dimensions.

The paper is organized as follows. Section \ref{sec:model} introduces the model of two identical interacting bosons subjected to a quasiperiodically modulated kicking sequence. Section \ref{sec:results} presents our numerical results and scaling analysis of the localization transition. Finally, Section \ref{sec:conclusion} summarizes our conclusions and discusses possible extensions.


\section{Model \label{sec:model}}

\medskip

We consider a system of $N$ identical $m$ mass bosons confined on a ring of length $L$ with periodic boundary conditions. These bosons experience point-like repulsive interactions of strength $g$ and are subjected to a kick potential of the form $\kappa(t)\cos(2\pi y_i/L)\sum_n \delta (t'-n\tau)$, with coordinates $y_i$, period $\tau$, and amplitude $\kappa(t)$. The Hamiltonian of this system $H = H_0 +  H_{\rm{kick}}$ is given in dimensionless units as 
\begin{equation}\label{eq_H_LL}
  H_0 = \sum_{i=1}^N \frac{p_i^2}{2}  +  c \, \sum_{i>j} \delta(x_i-x_j),
\end{equation}
the Lieb-Liniger Hamiltonian, which can be exactly solved with Bethe Ansatz \cite{LiebLiniger}, and
\begin{equation}\label{eq_H_K}
  H_{\rm{kick}}=H_K \sum_{n=-\infty}^{+\infty}\delta(t-n)\,,\quad H_K=\mathcal{K}(t) \sum_{i=1}^N\cos(x_i),
\end{equation}
the kick Hamiltonian. Our dimensionless system is obtained after rescaling time with $\tau$, positions with $\ell=L/2\pi$ and energy with $\varepsilon=m\ell^2/\tau^2$, where $c=g/(\varepsilon\ell)$, $\mathcal{K}(t)=\kappa(t)/(\tau\varepsilon)$, $t=t'/\tau$, and $x_i=y_i/\ell$. The momenta of the particles is given as $p_i=-i\hbar_e \partial / \partial x_i$, which satisfies the commutation relation $[x_j,p_j]=i \hbar_e$, with
$\hbar_e=\hbar/(\varepsilon\tau)$ as the effective Planck's constant. 

If $\mathcal{K}(t)$ remains constant, our model can be mapped to an Anderson model with dimensions equal to the particle number of the system \cite{Olsen2025}. However, with careful tailoring of the time dependence of $\mathcal{K}(t)$ one is able to realize a quasi-periodic quantum kicked rotor, changing the effective dimensionality of the system \cite{Casati1989b,Chabe2008,Madani2025}. In our work, we focus on two forms of $\mathcal{K}(t)$,    
\begin{equation}\label{eq_K_t}
  \mathcal{K}(t) = K(1 + \varepsilon \cos(\omega_2 t + \phi_2) ) 
\end{equation}
and 
\begin{equation}\label{eq_K_t2}
  \mathcal{K}(t) = K(1 + \varepsilon \cos(\omega_2 t + \phi_2) \cos(\omega_3 t + \phi_3) ). 
\end{equation}
To match the experiments \cite{Chabe2008}, we use $\omega_2 = 2 \pi \sqrt{5}$, $\omega_2 = 2 \pi \sqrt{13}$ and set $\phi_2=\phi_3=0$. The angular frequencies, $\omega_1$, $\omega_2$, $\hbar_e$, and $2 \pi$ must be incommensurate so that no unwanted resonances appear in the system \cite{SANTHANAM20221}. In the rest of the paper we use $\hbar_e=2.89$.

To study the dynamics of this system, it is convenient to work in the eigenbasis of the Lieb-Liniger Hamiltonian. Following Lieb and Liniger \cite{LiebLiniger}, this eigenbasis is obtained using a Bethe ansatz, giving eigenstates in the fundamental sector $x_1\le x_2\le ...\le x_N$ of the form
\begin{equation}\label{eq_psi}
  \Psi_{\{\lambda_j\}}(\{x_j\}) = \sum_{P\in S_N} A_P  \exp\Bigl(i \sum_{k=1}^N \lambda_{P(k)} x_k\Bigr),
\end{equation}
labeled by a set of $N$ rapidities $\lambda_i$. Here $S_N$ is the permutation group of $N$ elements and the $A_P$ coefficients are given by
\begin{equation}\label{eq_ap}
  A_P = \mathcal N (-1)^P  \prod_{1\le k < j \le N}[\lambda_{P(j)} - \lambda_{P(k)}-ic]
\end{equation}
with $\mathcal N$ the normalization constant and $(-1)^P$ the signature of permutation $P$. Rapidities must be real and satisfy
\begin{equation}\label{eq_bethe}
  \lambda_j = I_j - \frac{1}{\pi} \sum_{k=1}^N \textrm{arctan}\left(\frac{\lambda_j - \lambda_k}{c}\right),
\end{equation}
for distinct Bethe numbers $I_j$ of integer value for odd $N$, and half-integer for even $N$. The ground state is given by the lowest available set of Bethe numbers, for example $\vec I=(-1/2,+1/2)$ for $N=2$. Energy and momentum of Bethe states are found respectively as $E_{\vec \lambda} = \frac{\hbar_e^2}{2} \sum_{i=1}^N \lambda_i^2$ and $P_{\vec \lambda} = \hbar_e \sum_{i=1}^N \lambda_i$. From a stroboscopic point of view the dynamics of the system is obtained through repeat application of the time evolution operator on the wavefunction. Our system is a Floquet system, thus the time evolution is found using the Floquet equation
\begin{equation}
  \label{eq_Ufloquet}
  U=e^{-i H_0/\hbar_e} e^{-i H_K(t)/\hbar_e}.
\end{equation}


In this work we consider the case of $N=2$ particles, allowing us to conduct extensive numerical calculations. To determine the dynamics of the system numerically, we consider the matrix form of the Floquet equation. Its matrix elements reduce, as $e^{-i H_0/\hbar_e}$ is diagonal in this basis, to
\begin{equation}
  \label{eq_Ufloquet2}
M_{{\vec \lambda},{\vec \mu}}  = \langle{\vec \lambda} \vert e^{-iH_K(t)/\hbar_e} \vert {\vec \mu} \rangle,
\end{equation}
where $\vert {\vec \lambda} \rangle$ and  $\vert {\vec \mu} \rangle$ denote two Bethe eigenstates.
These matrix elements are computed analytically using \eqref{eq_psi} and their explicit values are given in App. \ref{sec:appendix}. To perform the time evolution numerically it is necessary to truncate the Hilbert space. We do this by restricting the Bethe numbers to the range $-N_s/2\le I_j\le N_s/2$ for $N_s = 101$ and $I_j$ a half-integer, resulting in a matrix size of $N_M=\binom{N_s+1}{2}=5151$.
Initial simulations of the rapidity distributions and energy evolution indicate that this range is sufficiently large to yield reliable results with minimal boundary effects. We also verify that if we consider larger $N_s$ physical variables remain largely unchanged. For the time evolution, the system is first initialized in the ground state, characterized by Bethe numbers $\vec I=(-1/2,1/2)$. It is then evolved numerically by repeatedly applying the Floquet operator, which must be recalculated at each time step because of the time dependence of the kicking potential. This process was completed for different forms of $K(t)$ and for selected values of $k$, $\varepsilon$, $c$, and $\hbar_e$. The total energy of the system is then straightforwardly found using
\begin{equation}
  \label{eq_EnergyNumeric}
 \langle E(t)\rangle = \sum_{\vec \lambda} |\alpha_{\vec \lambda}(t)|^2 E_{\vec\lambda}
\end{equation}
where $\alpha_{\vec \lambda}(t)$ is the amplitude of the many-body wave function in a given Bethe state $|\vec\lambda\rangle$. 

Whilst studying the dynamics of these systems, one problem we encounter is that various of the observables display large fluctuations during the time evolution. One way to reduce the effects of these fluctuations on our results is to average over the dynamically conserved quantity of the quasi-momentum $\beta$. Changing the quasi-momentum in the quantum kicked rotor system is equivalent to changing the disorder realization in the standard Anderson model \cite{Lemarie2009}. Quasi-momentum is introduced to our system simply by adding a quasi-momentum term to the energy of the Lieb-Liniger model similarly to what has been done in Ref.~\cite{Chicireanu2021}. In practice we average over fifty values of $\beta$ randomly chosen in $[0,1/2]$ with a uniform distribution.
For testing our numerics it is beneficial to understand the expected behavior in the limiting cases. In the non-interacting limit the system reduces to $N$ independent 1D quantum kicked rotors, each undergoing dynamical localization. In the infinite interaction limit one obtains an effective fermi exclusion principle, meaning a Bose-Fermi mapping can be applied to the system \cite{Girardeau}, resulting in localization properties quantitatively similar to the non-interacting case \cite{Vuatelet2021, RylandsRozenbaumGalitskiKonik}. We also know that for the standard two particle finite interacting case no transition from a localized to a delocalized phase is expected. The system remains localized and energy saturates for all values of the interaction. However, the final saturation energy depends on the interaction strength, reaching its minimum as both the non-interacting and infinite interaction limits are approached \cite{ChicireanuRancon}. 

\section{Results \label{sec:results}}

\medskip

We now present our numerical results together with the finite-time scaling analysis used to characterize the possible phase transition. Before discussing the data, we recall that in the orthogonal symmetry class, an Anderson transition is expected to occur only in dimensions $d\ge 3$ \cite{Abrahams1979,Evers2008}. In our model, the effective dimensionality is determined by the number of particles $N$ (for finite interactions) and by the number of additional driving frequencies $N_\omega$, $d=N+N_\omega$. This is a consequence of the fact that the addition of $N_\omega$ driving frequencies adds $N_\omega$ extra dimensions which are the same for each particle \cite{Casati1989b}.

Here we focus on the case of two interacting bosons ($N=2$) and consider $N_\omega=0,1,2$, which allows us to explore effective dimensions from two to four.  We also recall that the critical exponents are supposed to follow the Wegner's scaling law \eqref{eq:wegner}.

\begin{figure}
   \centering
    \includegraphics[width=0.45\textwidth]{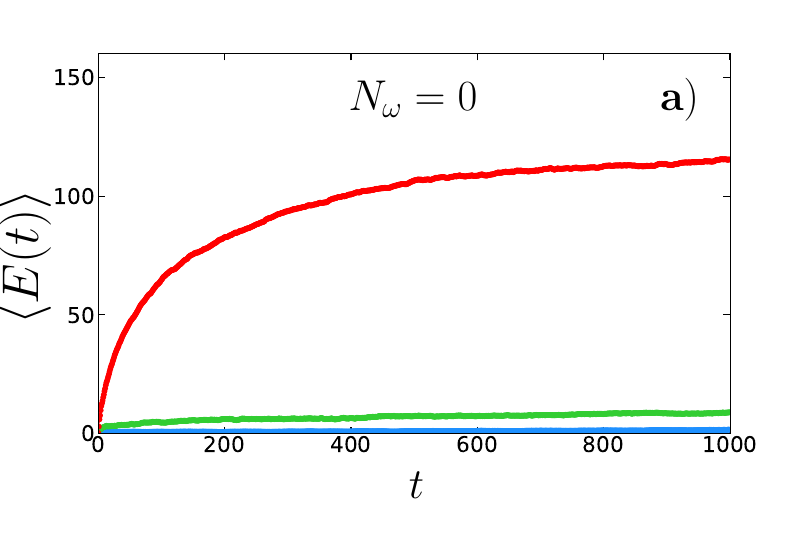}
    \includegraphics[width=0.45\textwidth]{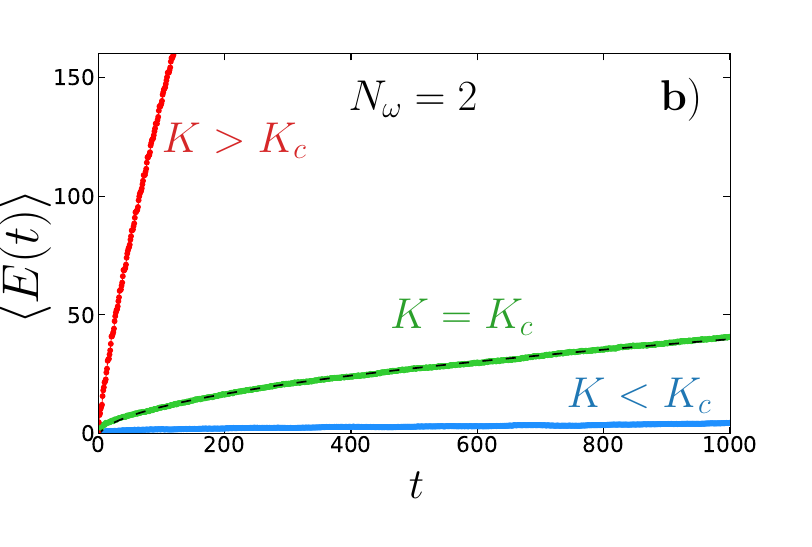}    
    \caption{Time evolution of the total energy of the two particles for different values of the stochasticity parameter $K$ at finite interaction strength $c=10$ and fixed $\varepsilon=0.3$ with a) $N_\omega=0$ and b) $N_\omega=2$. In the former case a), no transition is observed and energy always saturates with different plateau depending on $K$ (here $K/\hbar_e=0.7,\,1.3,\,2.3$) as expected from scaling theory of localization.  In the latter case b) a transition is observed. Close to the critical point $K/\hbar_e=1.3$ (green curve) anomalous diffusion is observed with exponent $1/2$ (black dashed line) consistent with four dimensional scaling. For $K=0.7\hbar_e<K_c$ (blue curve) the energy growth saturates as expected in the localized regime. Above the critical point $K=2.3\hbar_e>K_c$ (red curve) the dynamics is diffusive.}
    \label{fig:Energy_vs_time}
\end{figure}

Our numerical procedure consists in iterating the Floquet operator [Eq.~\eqref{eq_Ufloquet}] over many periods, starting from the ground state of the two-particle Lieb–Liniger model. From the resulting time-evolved states, we compute the average energy [Eq.~\eqref{eq_EnergyNumeric}] for different values of the stochasticity parameter $K$ at fixed (non-zero) interaction $c$ and modulation amplitude $\varepsilon$. Typical examples are shown in Fig.~\ref{fig:Energy_vs_time} for $N_\omega=0$ and $N_\omega=2$. In the latter case, a transition between localized and diffusive regimes is observed, as expected from the single-parameter scaling theory of localization \cite{Abrahams1979}, since $d=4$. For $K<K_c$, the energy saturates at long times, indicating localization. For $K>K_c$, it grows linearly with time, consistent with diffusive dynamics. At the critical point $K=K_c$, the growth becomes subdiffusive, following the scaling $\langle E(t)\rangle\sim t^{2/d}$. In this example, the exponent is $1/2$ and is consistent with our numerical data. Although not shown explicitly, the results for $N_\omega=1$ display the same qualitative behavior with a subdiffusive exponent $2/3$. No transition is observed for $N_\omega=0$, in agreement with the absence of an Anderson transition in two dimensions.

These observations, however, do not by themselves constitute proof of the existence (or absence) of a genuine phase transition, since our simulations are necessarily limited to finite evolution times. To establish the presence of a true second-order transition and to extract the corresponding critical exponents, we perform a finite-time scaling analysis \cite{Lemarie2009} and test whether all data collapse onto the universal scaling function predicted by localization theory. 

\begin{figure}
    \centering
    \includegraphics[width=0.45\textwidth]{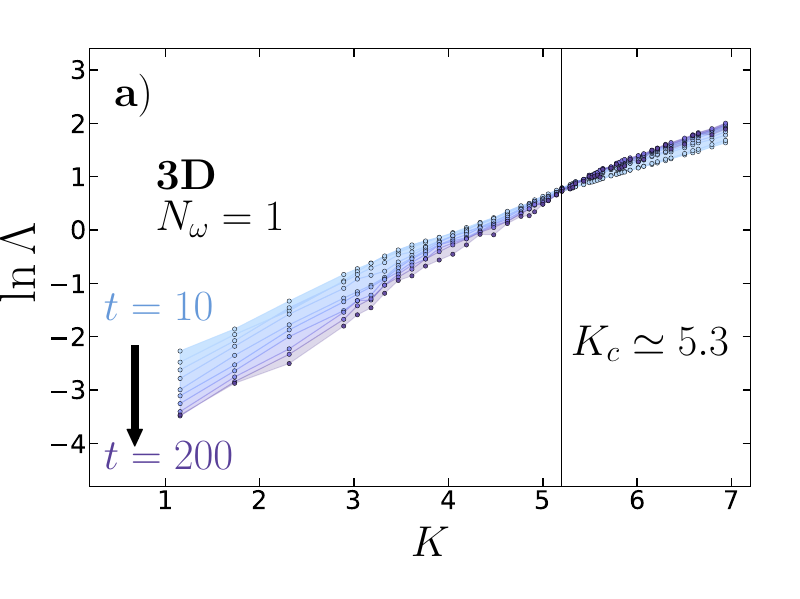}
    \includegraphics[width=0.45\textwidth]{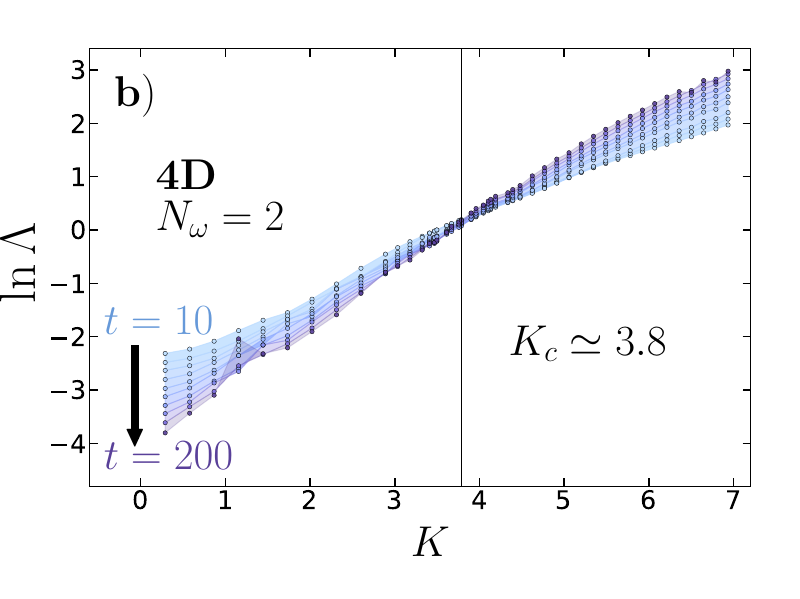}    
    \caption{The rescaled quantity $\ln\Lambda=\ln\left[\langle E(t)\rangle/t^{2/d}\right]$ as a function of $K$ for different times between $t=10$ and $t=200$. All curves intersect approximately at the critical point demonstrating the existence of a metal-insulator transition and allowing to locate the critical point $K_c$. Here $c = 10$, $\varepsilon = 0.3$ and $N_\omega=1$ for the left figure ($d=3$) and $N_\omega=2$ for the right one ($d=4$).}
    \label{fig:Kcrossing}
\end{figure}

The central idea of this approach is that, after an initial transient, the system’s dynamics becomes universal and can be rescaled using a single characteristic parameter, the correlation length $\xi$, which in our case has the dimension of momentum. All observables are then expected to collapse onto a single universal curve that depends only on the effective dimensionality $d$. To test this hypothesis, we rescale the energy dynamics according to the expected critical behavior $\langle E(t)\rangle \sim t^{2/d}$ \cite{Lemarie2009}, and define the dimensionless quantity
\begin{equation}
  \label{eq_LambdaDef}
  \Lambda(K,t) = \langle E(t)\rangle\, t^{-2/d} = f\!\left( \xi(K)\, t^{-1/d} \right),
\end{equation}
where $f$ is the universal scaling function.

If a phase transition occurs, the scaling function is expected to exhibit the following asymptotic behaviors. In the localized regime, $\langle E(t)\rangle \sim \xi^2$, so that $\Lambda \sim \xi^2 t^{-2/d}$, which vanishes at long times. In this case, $\xi$ can be interpreted as the localization length—or, more precisely, as a localization momentum scale. At the critical point, $\Lambda$ becomes time-independent, corresponding to scale invariance of the dynamics. In the diffusive regime, $\langle E(t)\rangle \sim D t$, where $D$ is the diffusion constant, and thus $\Lambda \sim D\, t^{1-2/d}$, with $\xi$ being inversely proportional to $D^{1/(d-2)}$. Close to the critical point, $\xi$ is expected to diverge with critical exponents $\nu$ (on the localized side) and $s$ (on the diffusive side), following Wegner’s relation~\eqref{eq:wegner}. As a direct consequence of this scaling form, no phase transition is expected to occur in two dimensions.

In order to demonstrate the existence of a phase transition and to accurately determine the critical point, we plot the scaling function $\Lambda$ (or equivalently its logarithm) as a function of the stochasticity parameter $K$ for various evolution times. As discussed above, $\Lambda$ decreases in the localized regime, increases in the diffusive regime, and remains constant at the critical point. This behavior is clearly observed in Fig.~\ref{fig:Kcrossing}. For $N_\omega = 1, 2$, corresponding respectively to effective dimensions $d = 3$ and $d = 4$, the curves at different times intersect at a well-defined value of $K$, identifying the critical point $K_c$.

We now aim to demonstrate the universality of the dynamics by showing that all data sets can be rescaled using a single parameter $\xi(K)$, from which the scaling function can be constructed. To this end, we employ the method developed in Ref.~\cite{Lemarie2009}. Specifically, we plot $\ln \Lambda$ as a function of $\ln(1/t^{1/d})$ for various values of $K$, at fixed interaction and modulation parameters $\varepsilon$. In this representation, rescaling with the correlation length corresponds to a horizontal shift by an amount $\ln \xi(K)$, allowing all data points to collapse onto a single universal scaling curve. The values of $\ln \xi(K)$ are determined by minimizing the distance between the corresponding values of $\ln[\xi(K)/t^{1/d}]$ for each $\ln \Lambda$.  

\begin{figure}
  \centering
   \includegraphics[width=0.45\textwidth]{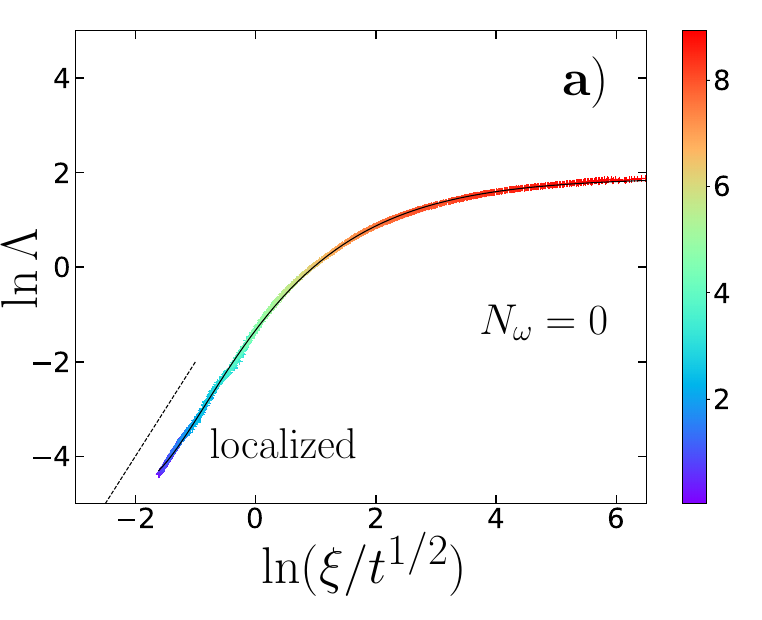}
   \includegraphics[width=0.45\textwidth]{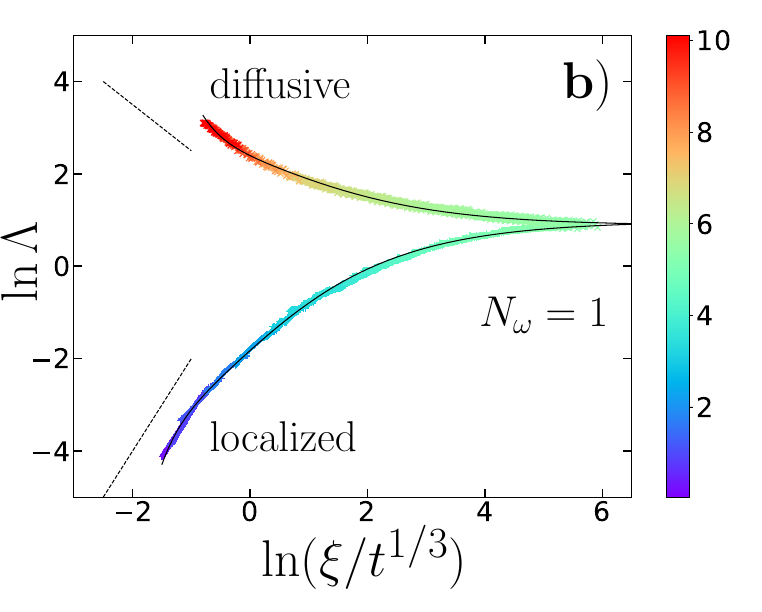}   
  \includegraphics[width=0.45\textwidth]{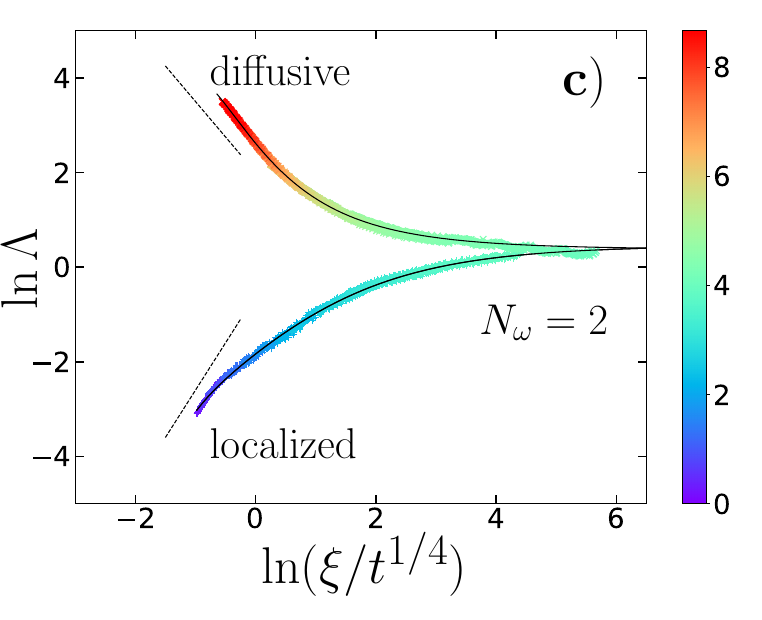}   
   \includegraphics[width=0.45\textwidth]{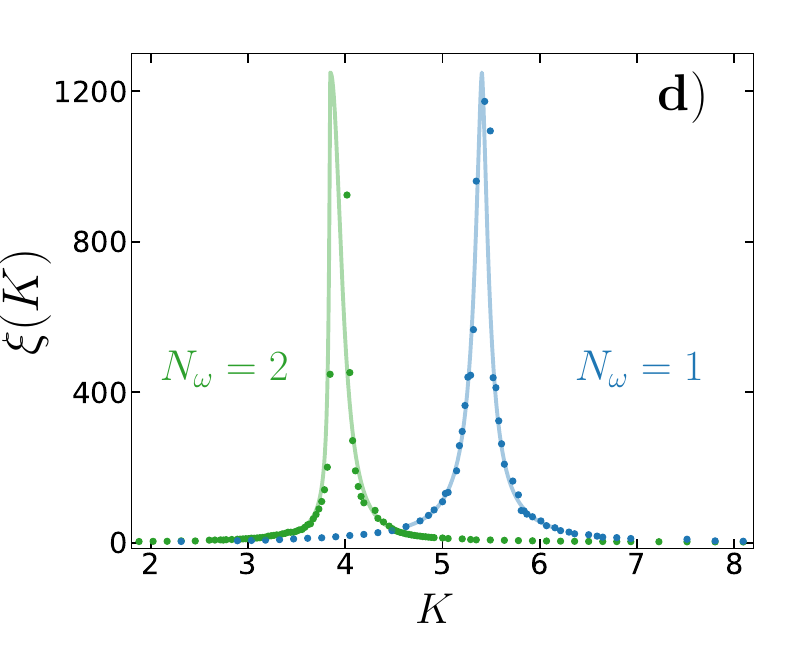}    
   \caption{
   Finite time scaling applied to the numerical results with $c=10$, $N=2$ and $\varepsilon=0.3$ for $N_\omega=0$ (a), $N_\omega=1$ (b) and $N_\omega=2$ (c). The time evolution of $\langle E \rangle$ is computed as a function of time, from 5 to 150 kicks, for several values of $K$ (see color bars). The scaling function $\ln\Lambda$, with $\Lambda=\langle E\rangle/t^{2/d}$, displays a lower branch (blue) associated with the localized regime and an upper branch (red) associated with the diffusive regime for $N_\omega\ge 1$ as a function of $\ln(\xi/t^{1/d})$. The continuous curves are fits using a Taylor expansion \eqref{eq:expansion_loglambda} of the scaling function up to fourth order with the corresponding critical exponents (see text). Dashed lines are the expected asymptotic behaviors. Panel (d) demonstrates the dependence of the scaling parameter $\xi(K)$ for the $N_\omega=1$ and $N_\omega=2$ cases. Here the continuous curves are fit either side of their respective critical points $K_c$ (see text), showing divergent behaviour around this point. From these fits we extract critical exponents of $\nu = 1.57 \pm 0.05$ and $s = 1.57 \pm 0.05$ for $N_\omega=1$, and $\nu = 1.17 \pm 0.04$ and $ s =  2.28 \pm 0.18 $ for $N_\omega=2$.}
  
\label{fig:ScalingFunctions}
\end{figure}

\begin{figure}
    \centering
    \includegraphics[width=0.45\textwidth]{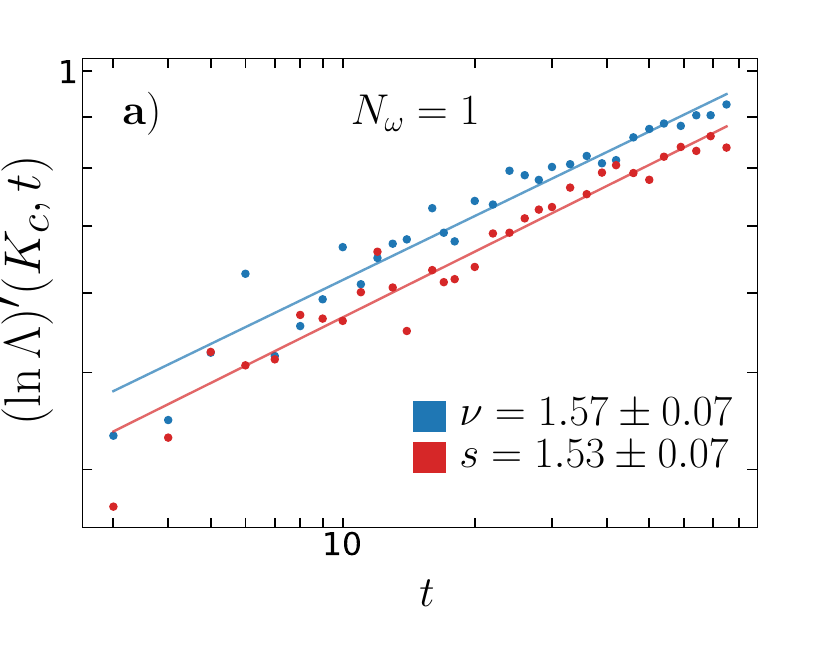}
    \includegraphics[width=0.45\textwidth]{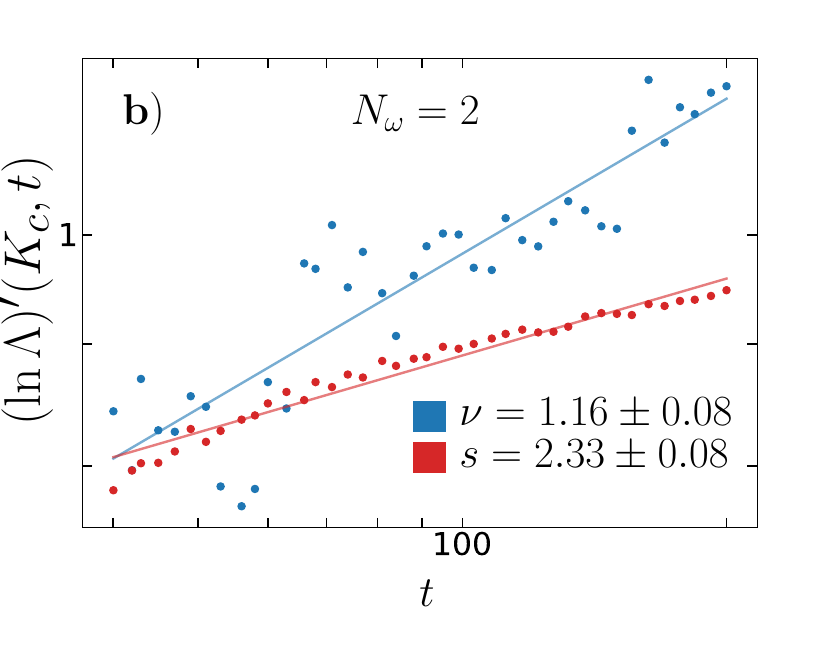}
    \caption{Determination of the critical exponents $\nu$ and $s$ by fitting $(\ln\Lambda)'(K_c,t)$ as function of $t$ in log-log scale. Here $c=10$, $\varepsilon=0.3$ and $N_\omega=1,2$ for a) and b).}
    \label{fig:critical_exponents}
\end{figure}

Our results are shown in Fig.~\ref{fig:ScalingFunctions} for the three cases under study. For $N_\omega = 0$, corresponding to an effective dimension $d = 2$, we recover a single localized branch of the scaling function, as expected from the absence of a phase transition in two dimensions. In contrast, Figs.~\ref{fig:ScalingFunctions}(b) and \ref{fig:ScalingFunctions}(c) display two distinct branches: the lower one associated with the localized phase and the upper one with the diffusive regime. This clearly reveals the existence of a metal–insulator transition. Moreover, the two branches merge at the critical point and exhibit the asymptotic behaviors discussed above. Fig.~\ref{fig:ScalingFunctions}(d) displays the scaling parameter $\xi(K)$ for $N_\omega=1, 2$, which diverges at each respective $K_c$. To obtain the power laws for these divergences, taking into account finite size effects, we fit our data using the function $\xi(K)^{-1} = \alpha \abs{K - K_c}^\nu + \beta$ below $K_c$, and $\xi(K)^{-1} = \alpha \abs{K - K_c}^s + \beta$ above $K_c$. From these fits we obtain power laws consistent with the relevant orthogonal Anderson universality classes, namely $\nu = 1.57 \pm 0.05$ and $s = 1.57 \pm 0.05$ for $N_\omega=1$, corresponding to an effective dimension of $d=3$, and $\nu = 1.17 \pm 0.04$ and $ s =  2.28 \pm 0.18 $ for $N_\omega=2$, corresponding to an effective dimension of $d=4$. In order to verify the values for these critical exponents, we use a more direct method as shown in Fig.~\ref{fig:critical_exponents} and described in the following. 
 
Having demonstrated the absence of a phase transition in our system for $N_\omega = 0$, and the existence of a metal–insulator transition for $N_\omega = 1, 2$, we now turn to the determination of the corresponding critical exponents. To this end, we assume an algebraic divergence of the correlation length in the vicinity of the critical point. In general, the power-law behavior differs on the two sides of the transition. As discussed above, $\xi(K) \sim (K_c - K)^{-\nu}$ in the localized phase, and $\xi(K) \sim (K - K_c)^{-s}$ in the diffusive phase. In order to extract the critical exponents $\nu$ and $s$, we expand the scaling function near the critical point as

\begin{equation}
\label{eq:expansion_loglambda}
\ln \Lambda(K,t) \simeq \ln \Lambda_c + 
\begin{cases}
\mathcal{C}_1^{<} (K_c - K)\, t^{1/d\nu} + \cdots, & \text{for } K < K_c, \\[3pt]
\mathcal{C}_1^{>} (K - K_c)\, t^{1/ds} + \cdots, & \text{for } K > K_c.
\end{cases}
\end{equation}

This implies that the slope of $\ln \Lambda(K)$ as a function of $K$ is discontinuous at $K_c$ (except for $d = 3$, where $s = \nu$) and follows power laws determined by the corresponding critical exponents. To determine these exponents, we measure the slopes in Fig.~\ref{fig:Kcrossing} as a function of time. The resulting data are displayed in a log–log plot in Fig.~\ref{fig:critical_exponents}. In the three-dimensional case, both exponents are equal, and a fit of Fig.~\ref{fig:critical_exponents}(a) yields $\nu = s = 1.59 \pm 0.04$, in excellent agreement with the state-of-the-art value for the three-dimensional Anderson transition in the orthogonal class, $\nu = 1.57 \pm 0.02$~\cite{Evers2008}. This result is also in agreement with the interaction induced Anderson transition for $N=3$ and $N_\omega=0$ reported in Ref.~\cite{Olsen2025} therefore supporting the universality of the transition in interacting kicked rotor systems. In the four-dimensional case ($N_\omega = 2$), shown in Fig.~\ref{fig:critical_exponents}(b), we obtain two distinct exponents that satisfy Wegner’s scaling law~\eqref{eq:wegner}, namely $\nu = 1.16 \pm 0.08$ and $s = 2.33 \pm 0.08$, again consistent with the most recent estimates of these critical exponents. For completeness, it would be important to compare this result with the $N=3$ and $N_\omega=1$ case as well as the $N=4$ and $N_\omega=0$ case, for which we also expect an Anderson transition in four dimensions. However, these cases are currently beyond our numerical capabilities. Finally, these critical exponent estimates are also in good agreement with our estimations from Fig.~\ref{fig:ScalingFunctions}(d).

The excellent agreement between our measured critical exponents and the well-established values for the Anderson transition, together with the observation of universal scaling behavior, provides a complete proof of concept that the system under study belongs to the orthogonal universality class of the Anderson transition in the effective dimension $d = N + N_\omega$. This result confirms that the two mechanisms used to generate additional dimensions combine in a simply additive manner, which was not a priori evident.

\begin{figure}
    \centering
    \includegraphics[width=0.45\textwidth]{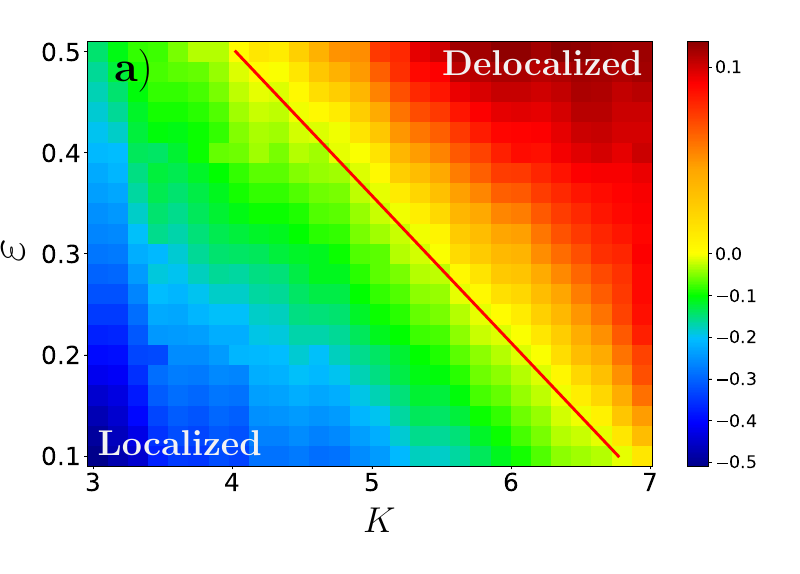}   
    \includegraphics[width=0.45\textwidth]{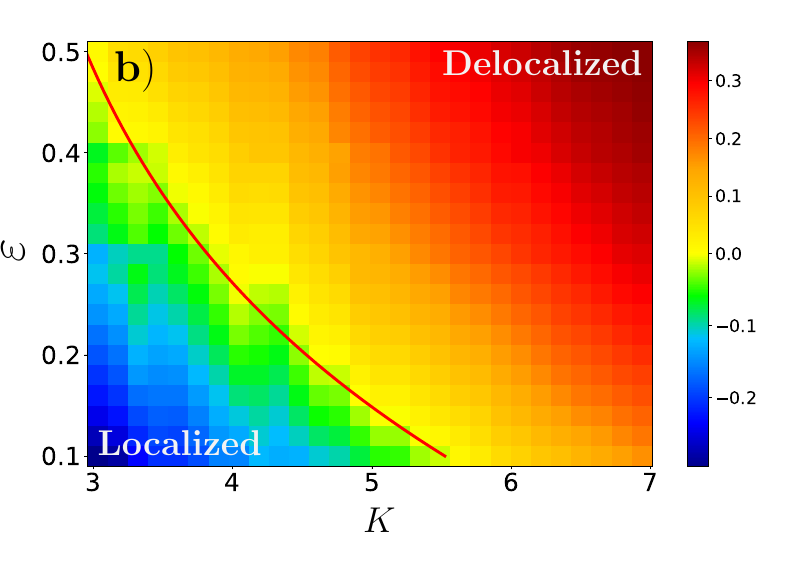}       
    \caption{Phase diagram of the dynamical phases in the $K$-$\varepsilon$ plane for two bosons with $c=10$. The color code corresponds to the derivative of $\ln\Lambda(t)$ at large time $t=170$ for a) $N_\omega=1$ and b) $N_\omega=2$. The red curve is the critical line where this derivative is zero.}
    \label{fig:phase_diagrams}
\end{figure}

Finally, we compute the phase diagram in the $(K, \varepsilon)$ plane for $N_\omega = 1$ and $N_\omega = 2$. The color scale in Fig.~\ref{fig:phase_diagrams} represents the time derivative of $\ln \Lambda$ at long times. Blue regions correspond to the localized phase, where this derivative is negative, while red regions indicate the diffusive regime, where it is positive. The solid line marks the critical boundary at which the derivative vanishes. We note that the color gradient observed near the transition reflects finite-time effects: in the limit of infinite evolution time, the transition becomes perfectly sharp, as expected for a genuine phase transition.

\section{Conclusion \label{sec:conclusion}}

\medskip

We have investigated the interplay of interactions and quasiperiodic driving in the kicked Lieb–Liniger model, establishing a simple and versatile framework to emulate Anderson localization in arbitrary dimensions. By combining two distinct mechanisms, namely interaction-induced coupling between particles and quasiperiodic modulation of the kicking strength, we have shown that synthetic dimensions can be engineered in a controlled manner. Focusing on the minimal case of two interacting bosons, we demonstrated that the system effectively realizes two-, three-, and four-dimensional Anderson models when driven with zero, one, and two additional incommensurate frequencies, respectively. Numerical simulations and finite-time scaling analysis confirm the existence of localization transitions characterized by critical exponents consistent with the orthogonal Anderson universality class.

Beyond its conceptual simplicity, this approach provides a powerful tool to explore Anderson localization and quantum criticality in regimes that are otherwise experimentally challenging to reach. Future directions include the development of experimental implementations using ultracold atomic gases or photonic systems, where both interactions and quasiperiodic driving can be precisely controlled. From a theoretical standpoint, our framework opens the way to investigate other symmetry classes, such as the unitary or symplectic ones, by appropriately modifying the driving protocol or introducing synthetic gauge fields. It also provides a natural platform to study properties beyond averaged quantities, including multifractality and nonergodic dynamics at criticality. Finally, while the formal mapping between the interacting Lieb-Liniger model and an effective Anderson model holds for arbitrary particle number $N$, our results explicitly demonstrate genuine Anderson localization only for $N=2$ and $N=3$ (shown in \cite{Olsen2025}). Whether increasing $N$ leads to a further growth of the effective dimensionality, or instead to a saturation of this mechanism, remains an open question that calls for future investigation.

\section*{Acknowledgments}
We would like to acknowledge helpful discussions with G. Aupetit-Diallo, R. Chicireanu, P. Devillard, G. Lemarié and A. Rançon. M. A. and P. V. are particularly grateful to D. Delande who introduced them to the physics of the kicked rotor. This work has benefited from the financial support of Agence Nationale de la Recherche under Grant No. ANR-21-CE47-0009 Quantum-SOPHA and Grant No. ANR-23-PETQ-0001 Dyn1D France 2030.

\appendix 

\medskip

\section{Calculation of the matrix elements in the Bethe basis}
\label{sec:appendix}

In this section, we recall the expressions of the eigenstates of the Hamiltonian (2) from the main text for the case of $N=2$ identical bosons \cite{LiebLiniger} and give the explicit expression of the matrix elements of the Floquet operator in this basis. Following the Bethe Ansatz, the wavefunction in each position sector is expressed as a superposition of plane waves with fixed energy. In the fundamental position sector $x_1<x_2$, the Bethe wavefunction takes the form 

\begin{equation}
    \Psi(x_1, x_2) =  A_{12} e^{i(\lambda_1 x_1 + \lambda_2 x_2)} + A_{21} e^{i(\lambda_1 x_2 + \lambda_2 x_1)} 
\end{equation}
where the positions $x_j$ are expressed in units of $\ell=L/(2\pi).$ 
The non-normalized amplitudes $A_{ij}$, which depend on the rapidities $\lambda_j$ and the interaction strength $c$, are determined by enforcing the cusp condition at $x_i=x_j$. This accounts for the discontinuity in the derivative due to the contact interactions \cite{LiebLiniger}. The amplitudes are given by

\begin{align*}
A_{12}(\lambda_1, \lambda_2, c) &= (\lambda_1 - \lambda_2 + ic)\\
A_{21}(\lambda_1, \lambda_2, c) &= (\lambda_1 - \lambda_2 - ic).
\end{align*}
Finally the rapidities $\lambda_j$ are obtained by imposing periodic boundary conditions, leading to the following system of coupled equations

\begin{equation}
\begin{split}
&\lambda_1=I_1 + \frac{1}{2\pi} \theta\left(\lambda_1 - \lambda_2\right), \\
&\lambda_2= I_2 + \frac{1}{2\pi} \theta\left(\lambda_1 -\lambda_2 \right)
\label{lambda_I}
\end{split}
\end{equation}
where $\theta(x) = -2 \arctan\left(x/c\right)$, and $I_1$ and $I_2$ are distinct relative half integers.

The non-trivial part of the Floquet operator to be evaluated is

\begin{equation}
M_{\vec\lambda,\vec\mu} = \langle \lambda_1, \lambda_2 | e^{-i K/\hbar_e (\cos x_1+\cos x_2)} | \mu_1, \mu_2 \rangle.
\end{equation}
This kind of matrix element has been obtained in Ref. \cite{Olsen2025} for arbitrary $N$. We therefore report the explicit results for $N=2$ only. It is expressed as
\begin{equation}
M_{\vec\lambda,\vec\mu} = \sum_{Q,Q'} \mathcal{A}_{Q,Q'} S_{Q,Q'},
\end{equation}
with
\begin{equation}
S_{Q,Q'} = 2 \sum_{\substack{m_1,m_2 }} (-i)^{ m_1+m_2} J_{m_1}(K/\hbar_e) J_{m_2}(K/\hbar_e)\, s_{Q,Q',\vec l},
\label{eq18}
\end{equation}
where
\begin{equation}
l_j = (\lambda_{Q(j)} - \mu_{Q'(j)}) + m_j,
\end{equation}
and $Q$ and $Q'$ are permutations of $S_2$. The explicit expression of $s_{Q,Q',\vec l}$ being 
\begin{equation}
s_{Q,Q',\vec l} =
-\frac{\,l_1 + e^{-i 2\pi (l_1 + l_2)}\, l_2 - e^{-i 2\pi l_2}\, (l_1 + l_2)\,}{\,l_1\, l_2\, (l_1 + l_2)\,}.
\end{equation}
When implemented numerically, the sum in Eq.~\eqref{eq18} must be truncated at finite $m_1$ and $m_2$. We have verified that our results do not change if we increase these cut off parameters.

We are able to eliminate artificial singularities in this expression by considering separately all the cases where
the denominator vanishes as follows:

\begin{equation}
s_{Q,Q',\vec l} =
\begin{cases}
 2\pi^2, & \text{if } l_1 = 0 \text{ and } l_2 = 0 \\[1em]
\dfrac{e^{-i 2\pi l_2}\left(1 - e^{i 2\pi l_2} + i 2\pi l_2\right)}{l_2^2}, & \text{if } l_1 = 0 \\[1em]
-\dfrac{-1 + e^{-i 2\pi l_1} + i 2\pi l_1}{l_1^2}, & \text{if } l_2 = 0 \\[1em]
\dfrac{1 - e^{i 2\pi l_1} + i 2\pi l_1}{l_1^2}, & \text{if } l_1 = -l_2 \\[1em]
-\dfrac{l_1 + e^{-i 2\pi (l_1 + l_2)} l_2 - e^{-i 2\pi l_2} (l_1 + l_2)}{l_1 l_2 (l_1 + l_2)}, & \text{otherwise.}
\end{cases}
\end{equation}

Finally, $K$ must be substituted by $\mathcal K(n)$ with $n$ the index of the period in order to take into account the quasi-periodic driving. In addition, those matrix elements have to be properly normalized. This is done by dividing the matrix elements by $\sqrt{\langle \vec \lambda|\vec\lambda\rangle \langle \vec \mu|\vec\mu\rangle}$ where $\langle \vec \lambda|\vec\lambda\rangle$ can be obtained from $M_{\vec\lambda,\vec\lambda}$ evaluated at $K=0$.

\printbibliography

\end{document}